\documentclass{article}
\usepackage{graphicx, multirow, listings, hyperref}
\usepackage{amsfonts, amsmath, amsthm}

\usepackage{color}

\newtheorem{theorem}{Theorem}

\begin{document}
\title{ Characterising epithelial tissues using persistent entropy \thanks{Partially supported  by MINECO, FEDER/UE under grant MTM2015-67072-P.
  Author names listed in alphabetical order.}
}
\author{N. Atienza, L.M. Escudero, M.J. Jimenez, M. Soriano-Trigueros}
%
%
\maketitle

\begin{abstract}
In this paper, we apply persistent entropy, a novel topological statistic, for characterization of images of epithelial tissues. We have found out that persistent entropy is able to summarize topological and geometric information encoded by $\alpha$-complexes and persistent homology
. After using some statistical tests, we can guarantee the existence of significant differences 
in the studied tissues. 

\end{abstract}
\section{Introduction}
Topological Data Analysis (TDA), originally, had its main motivation in the study of topological analysis of point cloud data. Nowadays is becoming a powerful tool for the study of shape of data, in its most general meaning. 
 The main tool used in TDA is {\em persistent homology} \cite{ELZ00,ZC05}, which studies the evolution of homology classes and their life-times (persistence) in an increasing nested sequence of spaces (that is called a filtration) and which is more informative that the homology class of the whole space.

Persistent homology has proved to be a useful tool in the study of shape analysis (in \cite{ferri}, some trends are described). Lately, most of the efforts in the area have been focused on developing a vector representation (obtained via persistent homology) that can be treated, afterwards, from machine learning point of view (see, for example, \cite{Adams17}). Such approaches have got the drawback of the need of large sample datasets, which is not usually the case in practice. For that reason, we are concerned with exploring the power of persistent entropy, which is a simple parameter that can be obtained from persistent homology and can be statistically studied.

Our main contribution is the application of persistent entropy as a tool for the characterization of biological tissues. 

 In the following Section, we describe the biological problem that motivated this work. Section \ref{background} recalls main concepts from TDA that will be used in the sequel. Section \ref{method} describes the particular way in which we make use of persistent homology concepts to topologically analyze the input data. Reports on the computations performed as well as some conclusions are collected in Section \ref{experiments}.
 We draw some ideas for future work in the last section.

\section{Motivation}
Epithelial cells are cells from animal embryos that will transform into one of the $4$ types of adult tissues: epithelia, connective tissue, nervous tissue and muscle tissue. Epithelia are packed tissues formed by tightly assembled cells. Their apical surfaces are similar to convex polygons forming a natural tessellation. Epithelial organization has been analyzed in various systems from a topological and biophysical perspective \cite{Farhadifar2007} \cite{Gibson2006} \cite{SanchezGutierrez2016} \cite{Shraiman2005}. These studies have been mainly based in the analysis of the polygon distribution of the tissues. A new approach has just been developed in \cite{epigraph},  were the authors have provided an image analysis tool(implemented in the open-access platform FIJI) to quantify epithelial organization based in computational geometry and graph theory concepts. 

In \cite{IWCIA17}, the authors first applied persistent homology, looking for other organizational traits that could improve the characterization of epithelia. Some initial experiments were described, working on two types of tissues: chick neuroepithelium (cNT) from chicken embryos and wing imaginal disc in the prepupal stage (dWP) from Drosophila. However, we would like also to compare the latter (dWP) with middle third instar wing discs (dWL), which are two proliferative stages separated by 24h development (and hence, with very similar organization). In this paper, we are concerned with the study of the discriminative ability of persistent entropy, discovering statistically significant differences between images of the three tissues. This work may open a door to the inclusion of persistent entropy as one more parameter to be taken into account in analysis tools like \cite{epigraph}.


\section{Background}\label{background}
	 The input used more frequently in topological data analysis is a point cloud in a metric space. In particular, we will work with points in the euclidean space $\mathbb{R}^2$ obtained from images. The procedure when applying persistent homology is the following. First, transform the information carried by the point cloud into a sequence of geometric figures called a filtration of simplicial complexes. Then, compute the homology (which intuitively can be seen as ``holes") for each simplicial complex and track how it evolves along the filtration. Finally, use a suitable way of representing the output and apply statistical methods to reach the conclusions.\\
	In this section we will define briefly these concepts. For a more concise introduction the reader could refer to \cite{computational}.
	
	\paragraph{Simplicial Complex.} A \emph{simplex} is the convex hull of a finite set of points $ \tau = \{ p_1, \ldots, p_n \}$. Any of the possible simplices contained in $\tau$ are called its \emph{faces}. A \emph{simplicial complex} $\mathcal{K}$ is formed by a set of simplices satisfying:
	\begin{enumerate}
		\item Every face of a simplex in $\mathcal{K}$ is also in $\mathcal{K}$.
		\item The intersection of two simplices in $\mathcal{K}$ is a face of both.
	\end{enumerate}

	\paragraph{Filtration.} A filtration is a finite increasing sequence of simplicial complexes
	\begin{equation*}\label{key}
	 \mathcal{K}_1 \subset \mathcal{K}_2 \ldots \subset \mathcal{K}_n = \mathcal{K}
	\end{equation*}
	It is commonly defined using a monotonic function $f: \mathcal{K} \rightarrow \mathbb{R}$ by which we mean that for $\delta, \tau \in \mathcal{K}$, $f(\delta) \leq f(\tau)$ if $\delta \subset \tau$. In this way, if $a_1 \leq \ldots \leq a_n$ are the function values of the simplices in $\mathcal{K}$, then $\mathcal{K}_i = f^{-1}(-\infty, a_i]$.

	\paragraph{Persistent Homology}
	 The inclusion $\mathcal{K}_i \subset \mathcal{K}_{i+1}$ induces a linear map between vector spaces $H_*(K_i) \rightarrow H_n(K_{i+1})$, where $H_n$ is the homology of dimension $n$. Intuitively when a homology class disappears (i.e., it is in $K_i$ but not in $K_{i+1}$ for some $i$), 
	 we say that it dies at time $i$. When a homology class appears by the first time 
	(i.e., it is in $K_i$ but not in $K_{i-1}$ for some $i$), 
	we say that it has been born at time $i$.
	
	\paragraph{Barcodes} The fact that an independent homology class is born at time $i$ 
	and dies at $j$ can be represented by an interval $ ((i,j)$.  Then, the output of persistent homology can be represented as a multiset 
	$\{ (i,j) \}$, where $(i,j)$ are birth-death values of arising homology classes. This is usually represented using barcodes as in Figure \ref{background}.

	    \begin{figure}[ht]
		\begin{center}
			\begin{tabular}{ |c|| c| }
				\hline\multicolumn{2}{|c|}
				{\hspace{0.1cm}
					\includegraphics [scale=0.6] {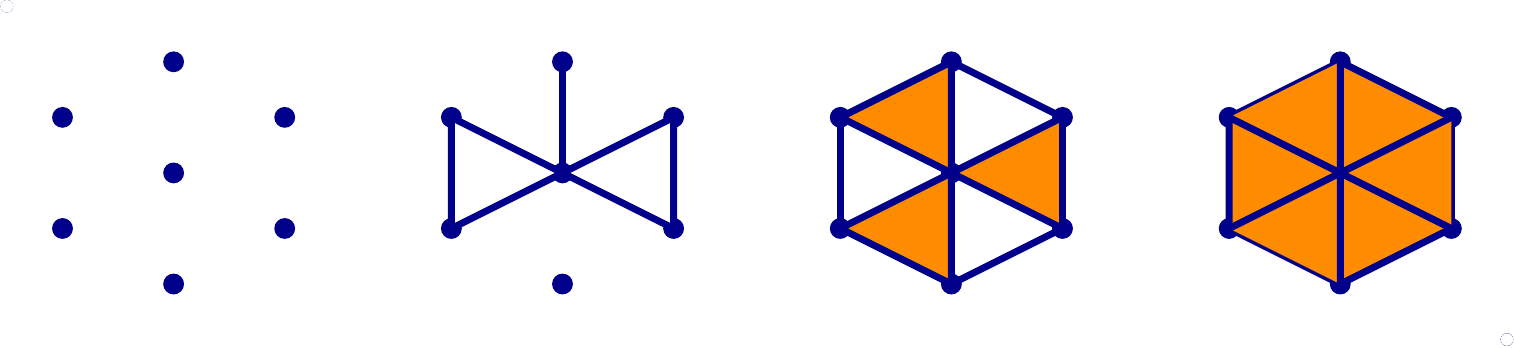}
					\hspace{0.45cm}}
				\\
				\hline
				\hline
				\begin{minipage}{.4\textwidth}
					\vspace{0.3 cm}
					\includegraphics [scale=0.35] {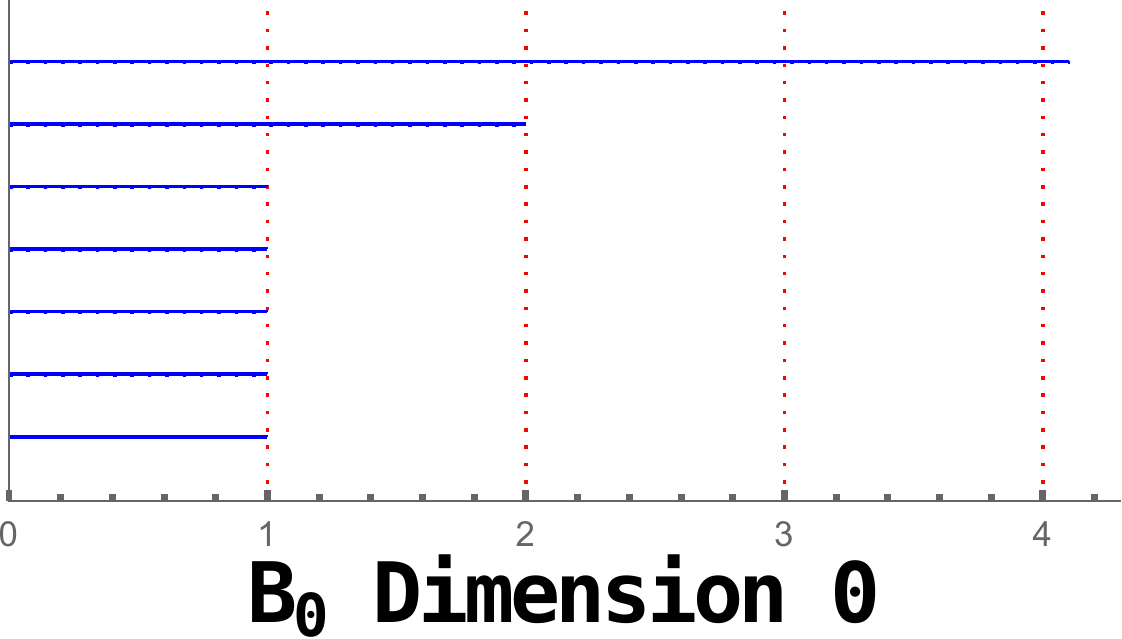}
				\end{minipage}
				&
				\begin{minipage}{.4\textwidth}
					\vspace{0.3 cm}
					\includegraphics [scale=0.35] {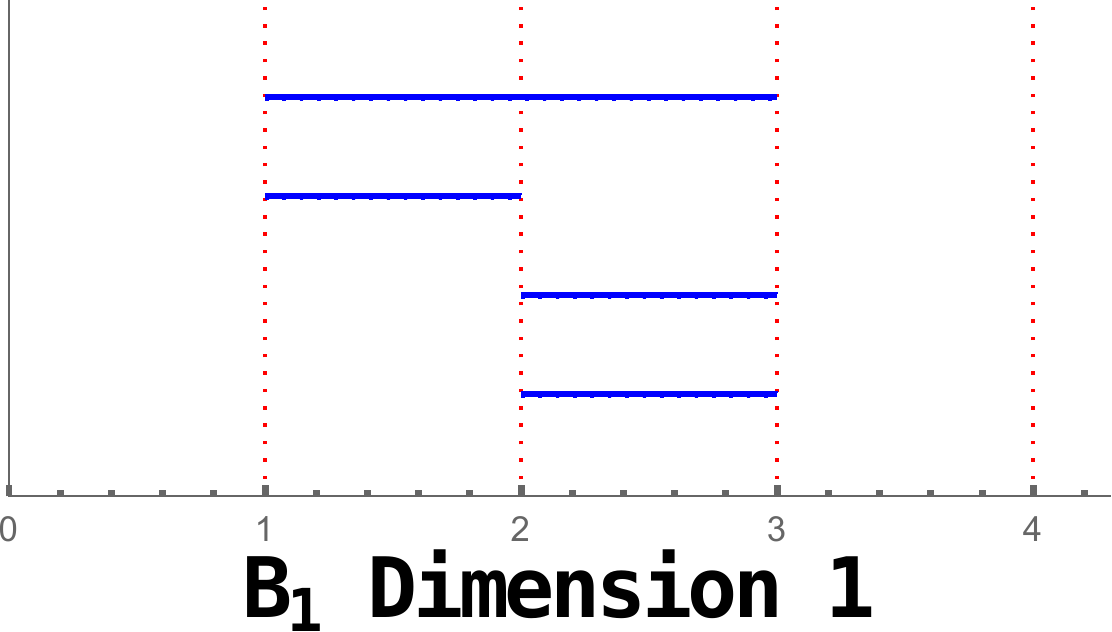}
				\end{minipage}
				\\
				\hline
			\end{tabular}
		\end{center}
		\caption{ Top: example of a filtration $\mathcal{K}$. Bottom: barcodes representing connected components ($0$-th dimensional Betti numbers) and cycles ($1$-th dimensional).}\label{figure:fig_bar}
	\end{figure}
	
	\paragraph{Persistent Entropy.}
	Persistence barcodes represent reliably the persistent homology output that could be treated statistically. However, the statistical tools used should be stable and robust to noise in the input data. 
	Under this premise, we make use of  \emph{persistent entropy} \cite{chintakunta2015entropy}, which can be defined as an adaptation of Shannon entropy to this context.
	
	Represent the multiset of birth and death using pairs $\{(x_i, y_i)\}_{i=1}^n$ (note that there might be repeated pairs). Consider the length of each of them $\ell_i = y_i - x_i$ and the total length $L = \ell_1 + \ldots + \ell_n$. Then, the persistent entropy of a barcode is the value
	\begin{equation*}
		E = \sum_{i=1}^n \frac{ \ell_i}{L} \log( \frac{ \ell_i}{L}).
	\end{equation*}
	The maximum possible value of persistent entropy is $\log(n)$ and is reached when all intervals have the same length. The minimum value is $0$ and coincides with the case $n=1$. In general, the greater the number of intervals is and the more homogeneous they are, the greater the persistent entropy is.
	
	The following result guarantees that persistent entropy is robust to noise, the proof appears in \cite{stability}. Before, we need some notation remarks: consider two barcodes $B_1$ and $B_2$ given by $B_j = \{(x_i^j, y_i^j)\}_{i=1\ldots n_j}$. The lengths of the bars are $\ell_i^j = y_i^j - x_i^j$ and their total length $L^j = \ell_1^j + \ldots \ell_{n_j}^j$. We denote the bottleneck distance for barcodes as $d_\infty$.
	
	\begin{theorem}\label{main_theo1}
	Let $\mathcal{K}$ be a simplicial complex  and let $f_1,f_2 : \mathcal{K} \rightarrow \mathbb{R}$ be two monotonic functions, $B_1$,$B_2$ their corresponding barcodes and $n_{max} = \max \{ n_1,n_2\}$. Then, if  $d_\infty(B_1,B_2) \leq \frac{1}{8} \frac{\max\{L^1, L^2 \}}{n_{\max}} $ and $||f_1 - f_2|| \leq \delta$, we have: 
		\begin{equation*}
			 |E(B_1) - E(B_2)| \leq \frac{4 \delta n_{\max}}{\max\{L^1, L^2 \}} \left[ \log(n_{\max})  - \log \left( \frac{4 \delta n_{\max}}{\max\{L^1, L^2 \}} \right) \right].\nonumber
		\end{equation*}
	\end{theorem}
	
	In other words, this theorem implies that if a maximum number of bars and a minimum length are fixed, then persistent entropy is uniformly continuous respect to the maximum norm of filter functions.

\section{Methodology}\label{method}

	In this section we will explain the steps involved in the method developed in this paper: 
	\begin{enumerate}
	    \item Normalize each image so that they all have the same number of cells.
		\item Consider the point cloud given by the centroids of the cells.
		\item Construct a simplicial complex called Delaunay Triangulation and a filtration on it, called the $\alpha$-complex, from the point cloud.
		\item Compute its persistent homology and persistent entropy.
		\item Perform a statistical study and analyze the results.
	\end{enumerate}

The input is an image with 1024x1024 pixels. This image is a gray scale image in which each segmented region corresponding to a cell has been labeled with an ID number and pixels on the boundary of cells are labeled by $0$. Now, we further develop some steps in the process:\\

As for Step $1$, an important drawback when using persistent entropy is that the number of cells affects its value. Then, if we want to measure  topological features using this parameter, we need to have the same number of cells for each sample. 
    This way, we have designed the next algorithm:\\

\begin{lstlisting}[escapeinside={*}{*}]
*\textbf{Input:} $n\in\mathbb{N}$ and $ M\in(\mathbb{N}^0)^{1024\times 1024}$.\\
\textbf{Output:} A set $\mathcal{C}$ of $n$ cells.*
*$\mathcal{C} := \emptyset$*
*$x=y=512$*
if *$M(x,y) \neq 0$*
    *$\mathcal{C}:= \{M(x,y)\}$*
i = 0
while  #*$\mathcal{C} < n$*
    i = i+1
    Repeat i times
        if #*$\mathcal{C} < n$*
            *$y := y + (-1)^{i+1}$*
            if *$M(x,y) \neq 0$ and  $M(x,y) \notin \mathcal{C}$*
               then *$\mathcal{C}:=\mathcal{C}\cup \{M(x,y)\}$*
    Repeat i times
        if #*$\mathcal{C} < n$*
            *$x := x + (-1)^{i}$*
            if *$M(x,y) \neq 0$  and $M(x,y) \notin \mathcal{C}$* 
               then *$\mathcal{C}=\mathcal{C}\cup \{M(x,y)\}$*
\end{lstlisting}

\begin{figure}[h!]
	\centering
	\begin{tabular}{c}
		\includegraphics[scale=0.9]{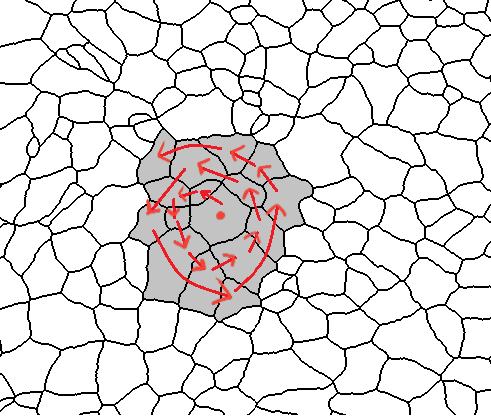}\\\\
		\includegraphics[scale=0.45]{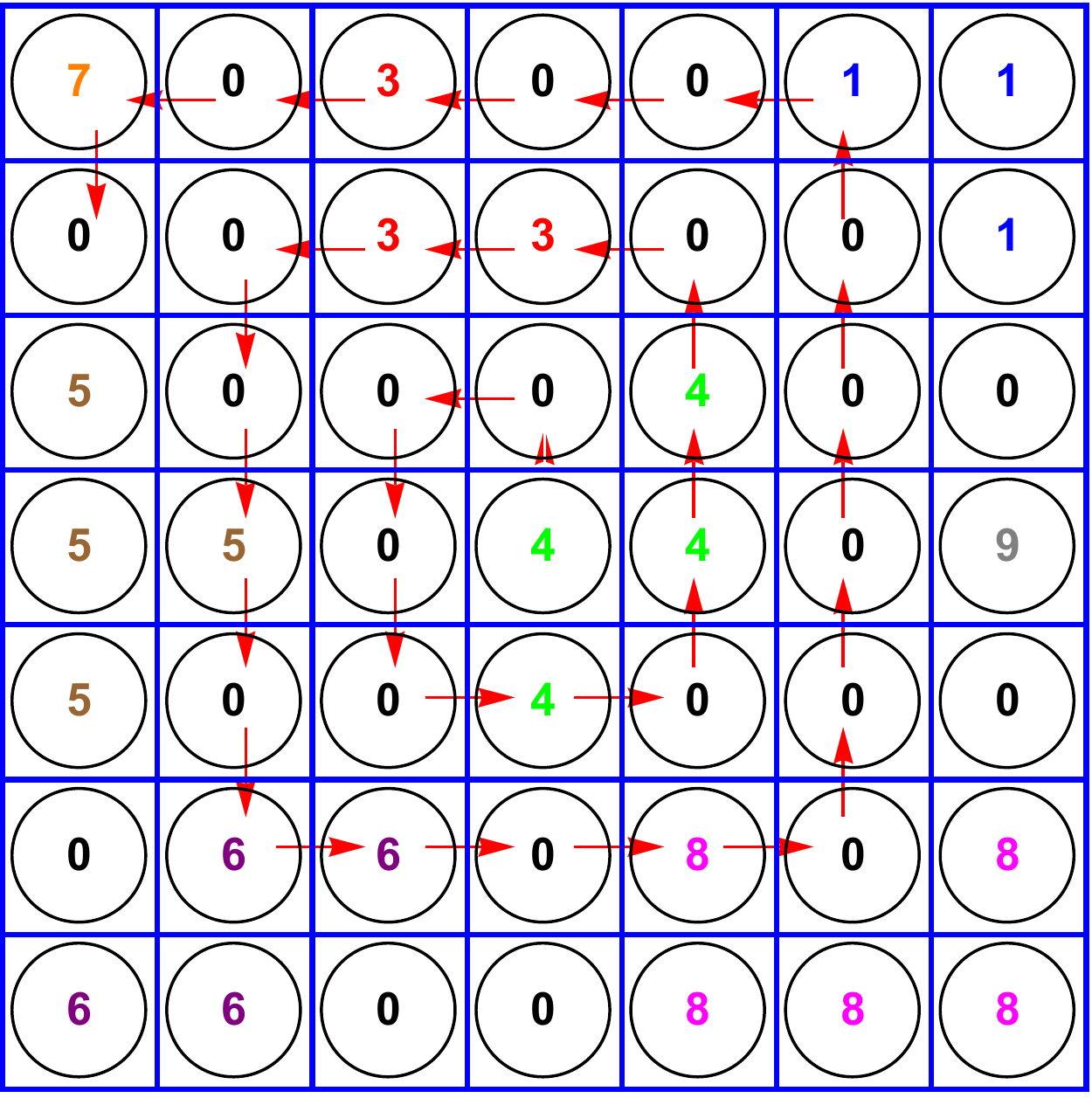}
	\end{tabular}
	\caption{Top picture illustrates the intuition behind the algorithm to restrict to a proper number of cells. Bottom: flow of the process at pixel level.}\label{espiral}
\end{figure}

Figure \ref{espiral} (bottom) shows a simple example in which, taking as input $n=7$ and the depicted pixel values, the output set of cells would be $\mathcal{C}=\{4,3,5,6,8,1,7\}$.

Regarding Step $3$, the main difficulty of this methodology is finding the proper filtration to distinguish the cell tissues. We have chosen $\alpha$-complex because it represents a good  approximation of the geometry of the cells. Nevertheless, this filtration may be improved as explained in Section \ref{end}. Here, we recall the basic concepts involved:

\emph{Voronoi Diagram}. A Voronoi Diagram is a partitioning of the plane depending on a set of vertices. For each vertex $v_i$ we define the function $f_i(x) = d(v_i,x)$ and a region given by 
\begin{equation}
	V_i = \{ x \; | \; f_i(x) \leq f_j(x) \quad \forall j\}.
\end{equation}

\begin{figure}[h!]
    \centering
    \includegraphics[scale=0.35]{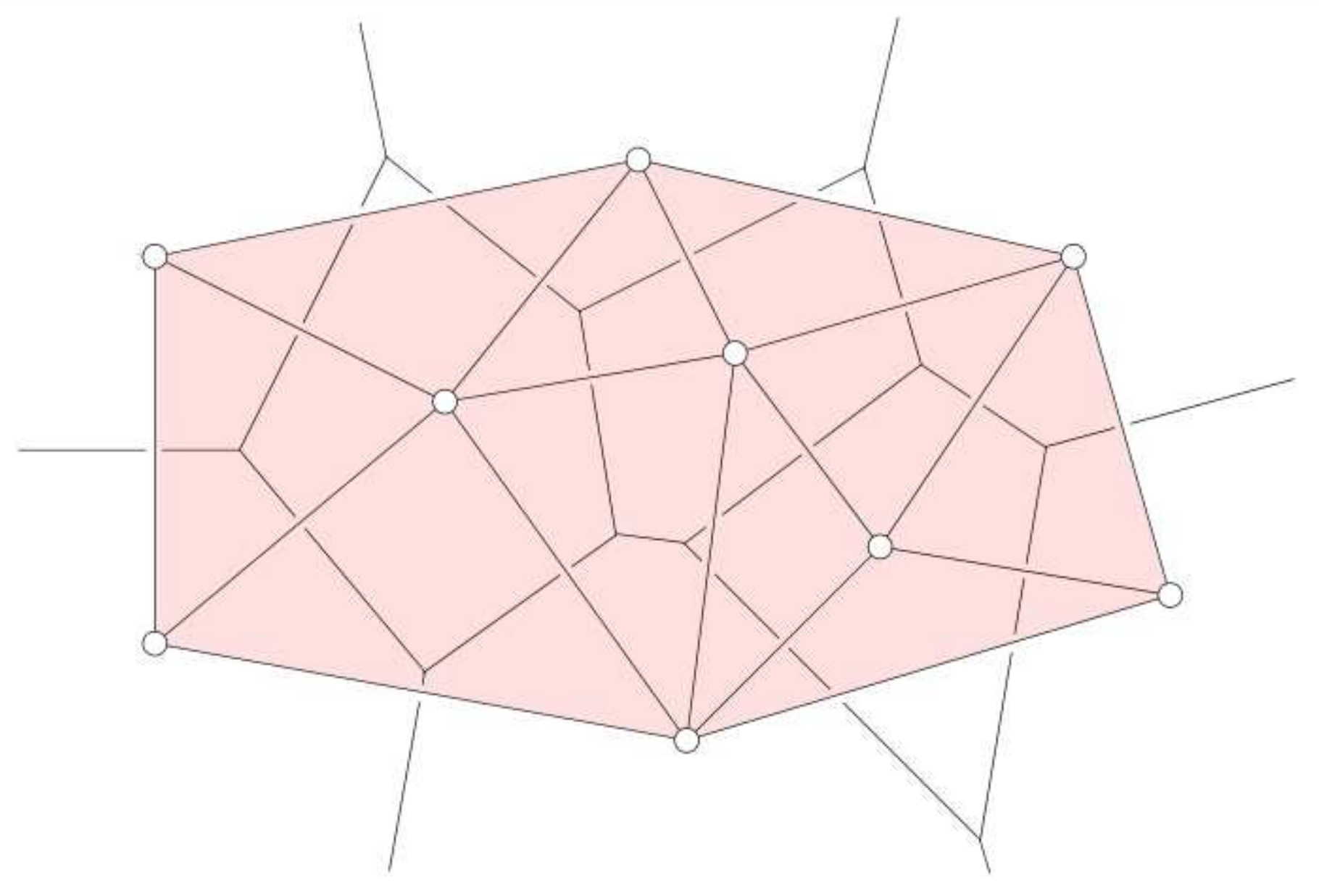} 
    \caption{Example of Dealunay Triangulation appearing in \cite{computational}}
    \label{fig:dealunay}
\end{figure}
\emph{Dealunay Triangulation}. A Dealaunay Triangulation is a simplicial complex which can be constructed from a Voronoi Diagram. An edge joins two vertices if the intersection of their Voronoi regions is not empty. The $2$-simplices are formed when three points have all possible edges between them. If the points are in general position in the plane, there are not simplices of greater dimension.

\begin{figure}

 	\begin{center}
	\includegraphics[scale=0.35]{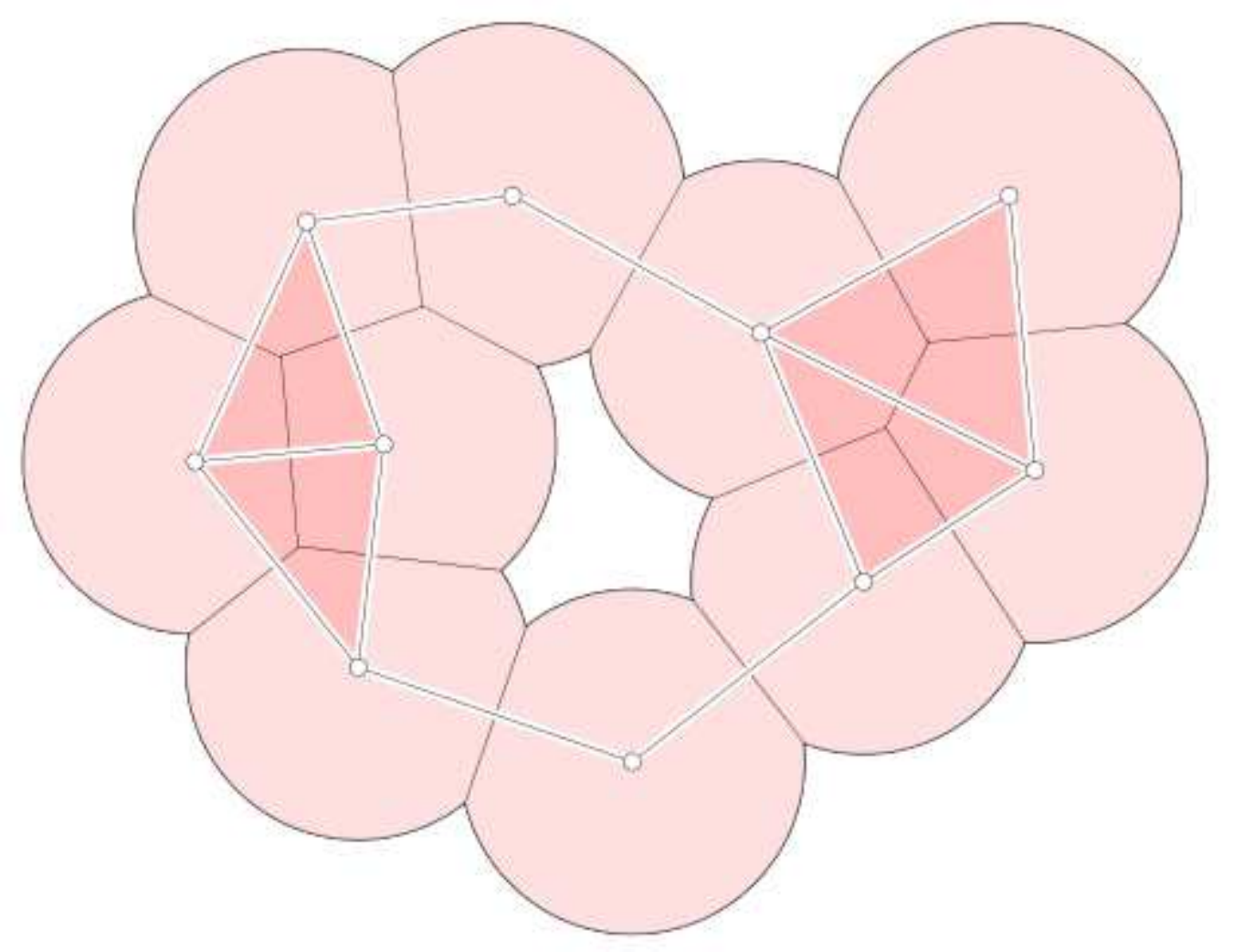}
 	\end{center}
  	\caption{Example of an Alpha Complex for a fixed $r$ appearing in \cite{computational}.}
 	\label{Alpha}
 \end{figure}

\emph{Alpha Complex}. We will use a filtration on the Delaunay Triangulation called $\alpha$-complex. Define $B_r^i$ as the ball of center $u_i$ and radius $r$. For each $r$, consider the region $U_r^i = B_r^i \cap V_i$ and define the simplicial complex $\mathcal{K}_r$ with simplices
\begin{equation}
	\tau = [ u_0 \ldots u_k ]\in \mathcal{K}_r \Leftrightarrow U_r^i \cap U_r^j \neq \emptyset \quad i,j = 0\ldots k.
\end{equation}
When $r$ is big enough, $U_r^i = V_i$ and the simplicial complex is just the Delaunay Triangulation. In particular, if the points are in general position in the plane, $n\leq 2$. See Figure~\ref{Alpha} for a picture.\\

	More specificaly, our filtration is computed as follows: we take the centroids of the cells as a set of vertices and compute their Voronoi Diagram. Using this structure, we construct the $\alpha$-complex filtration on top of this  Delaunay Triangulation. 
	
	Finally, we compute persistent entropy from persistence barcodes and make a statistical analysis of the results, what will be detailed in next section.

\section{Experiments and Results}\label{experiments}
Our database consists on $16$ images of chick neuroepithelium (cNT), $15$ images of Drosophila
wing imaginal disc from the third instar larva (dWL) and $13$ from the prepual state (dWP). All the images are obtained in a standard way. More information about the database is available in \cite{cell_images}. 

The number of cells of each image is shown in Table \ref{tab3}. As explained before, We need to fix a number of cells before the experiment. We choose $n=400$ since this is, approximately, the minimum number of cells appearing in the different samples. 

  \begin{table}
  	\begin{center}
  		\caption{Number of cells in each picture.}
  		\label{tab3}
  		\resizebox{\columnwidth}{!}{%
  		\begin{tabular}{|c|c|c|c|c|c|c|c|c|c|c|c|c|c|c|c|c|c|c|}
  			\hline
  			& 1 & 2 & 3 & 4 & 5 & 6 & 7 & 8 & 9 & 10 & 11 & 12 & 13 & 14 & 15 & 16 \\
  			\hline
	  		cNT & 666 & 661 & 565 & 573 & 669 & 532 & 419 & 592 & 743 & 527 & 594 & 473 & 704 & 747 & 469 & 834 \\
	  		\hline
	  		dWL & 426 & 555 & 491 & 522 & 510 & 936 & 890 & 789 & 977 & 913 & 604 & 835 & 785 & 747 & 622 & \\
	  		\hline
	  		dWP & 748 & 805 & 566 & 414 & 454 & 654 & 751 & 713 & 503 & 430 &  516 & 413 & 455 &  &  & \\
	  		\hline
  		\end{tabular}
	  	}
  	\end{center}
  \end{table}

\begin{figure}
	\begin{center}
		\begin{tabular}{c c}
		\includegraphics[scale=0.13]{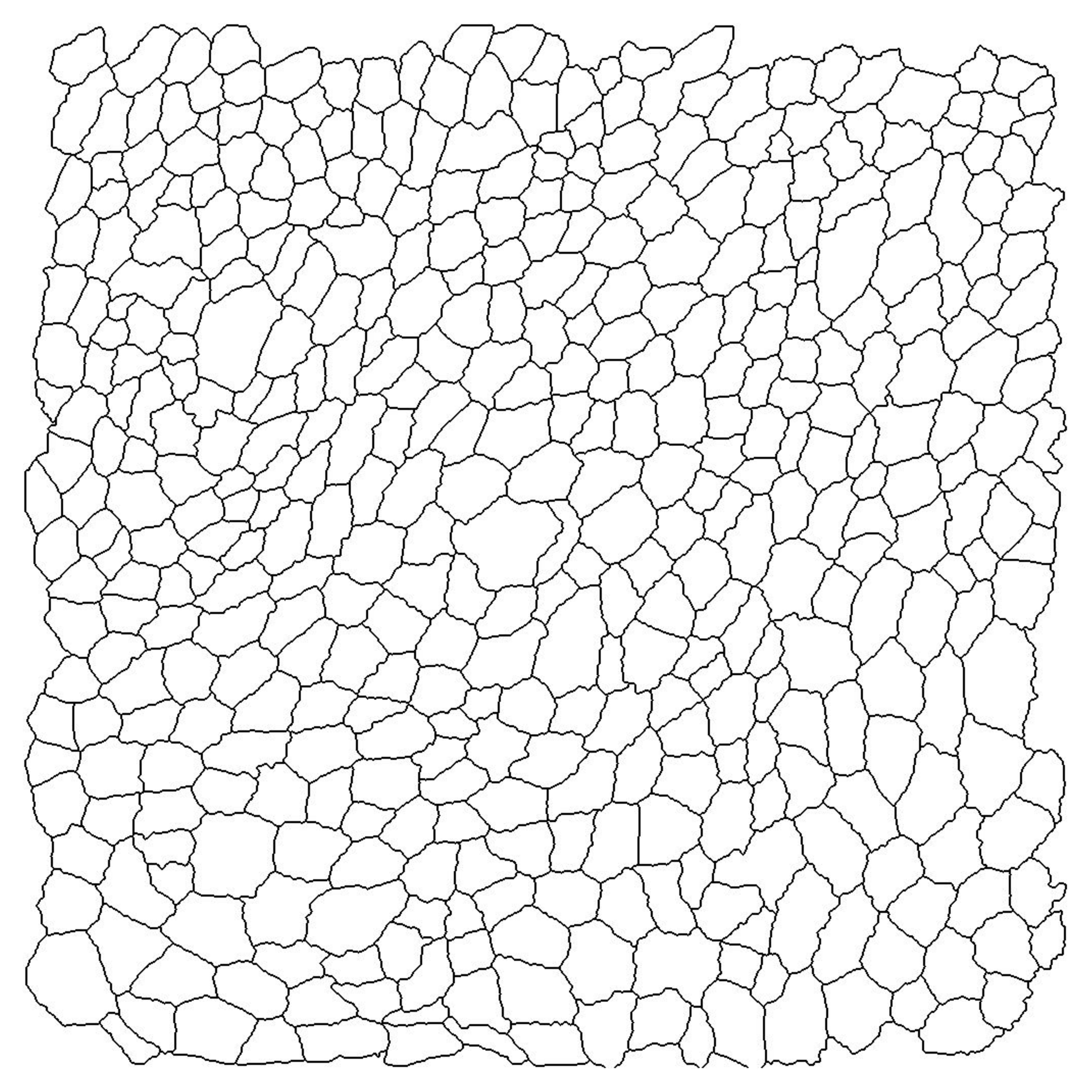} & \includegraphics[scale=0.13]{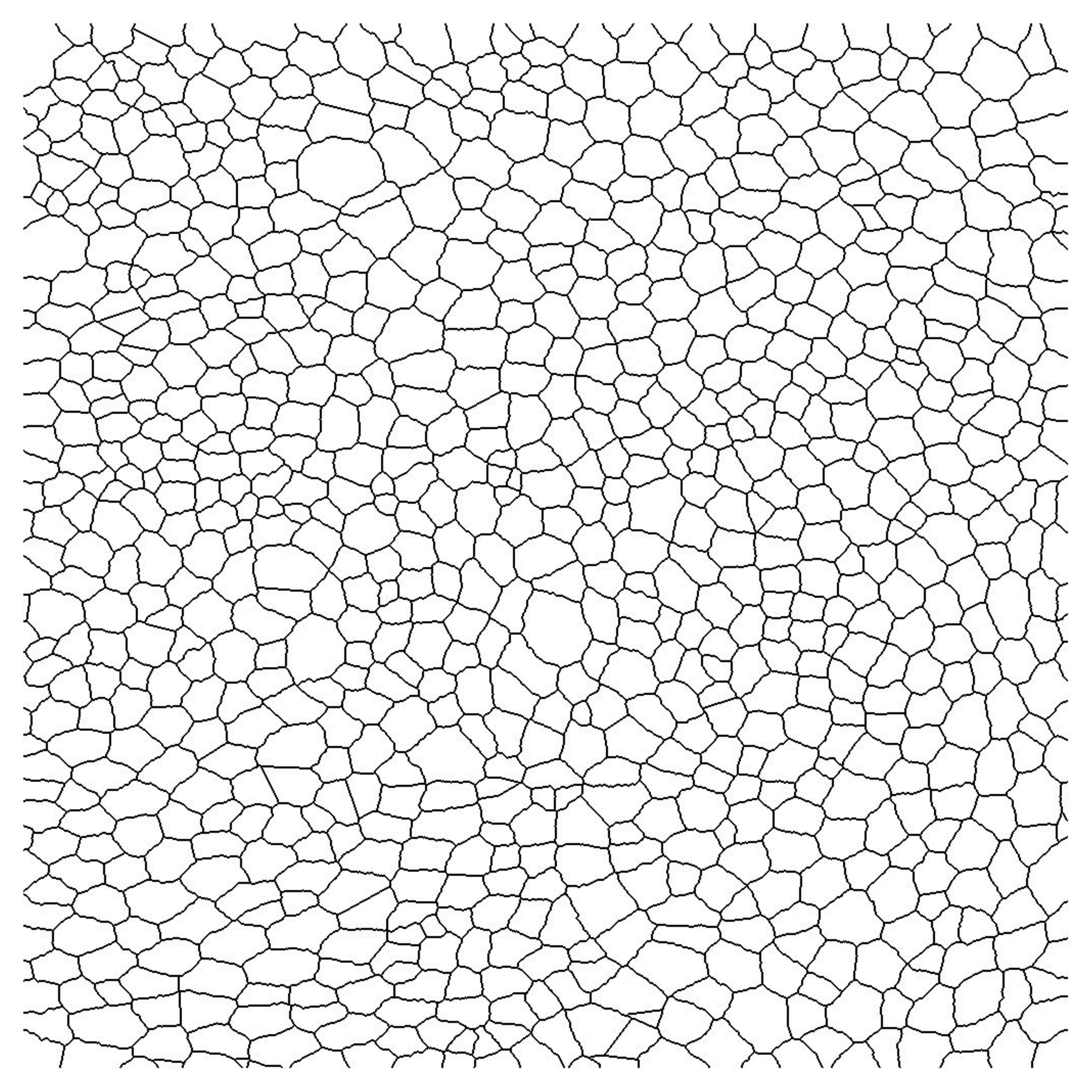} \\ \includegraphics[scale=0.35]{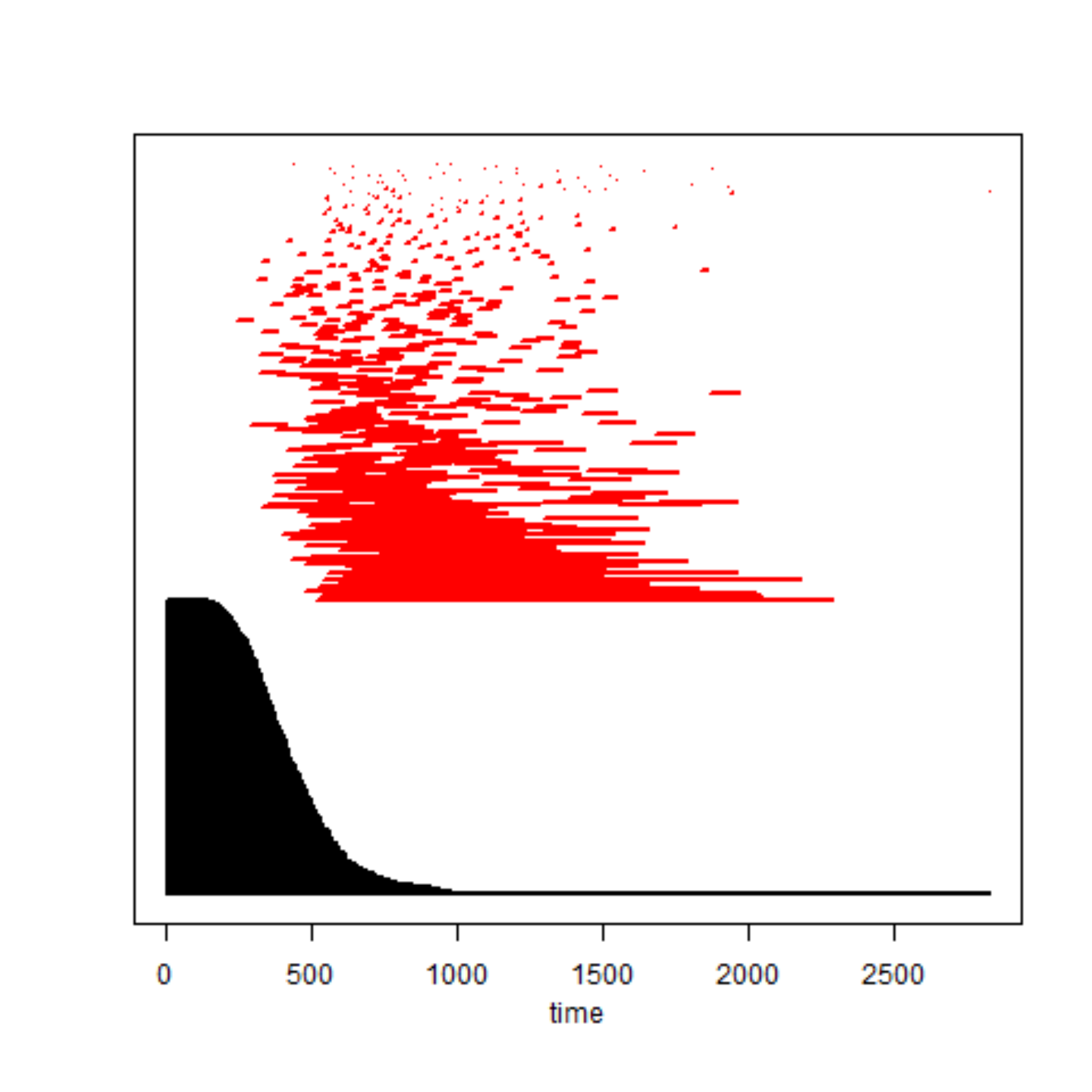}  & \includegraphics[scale=0.35]{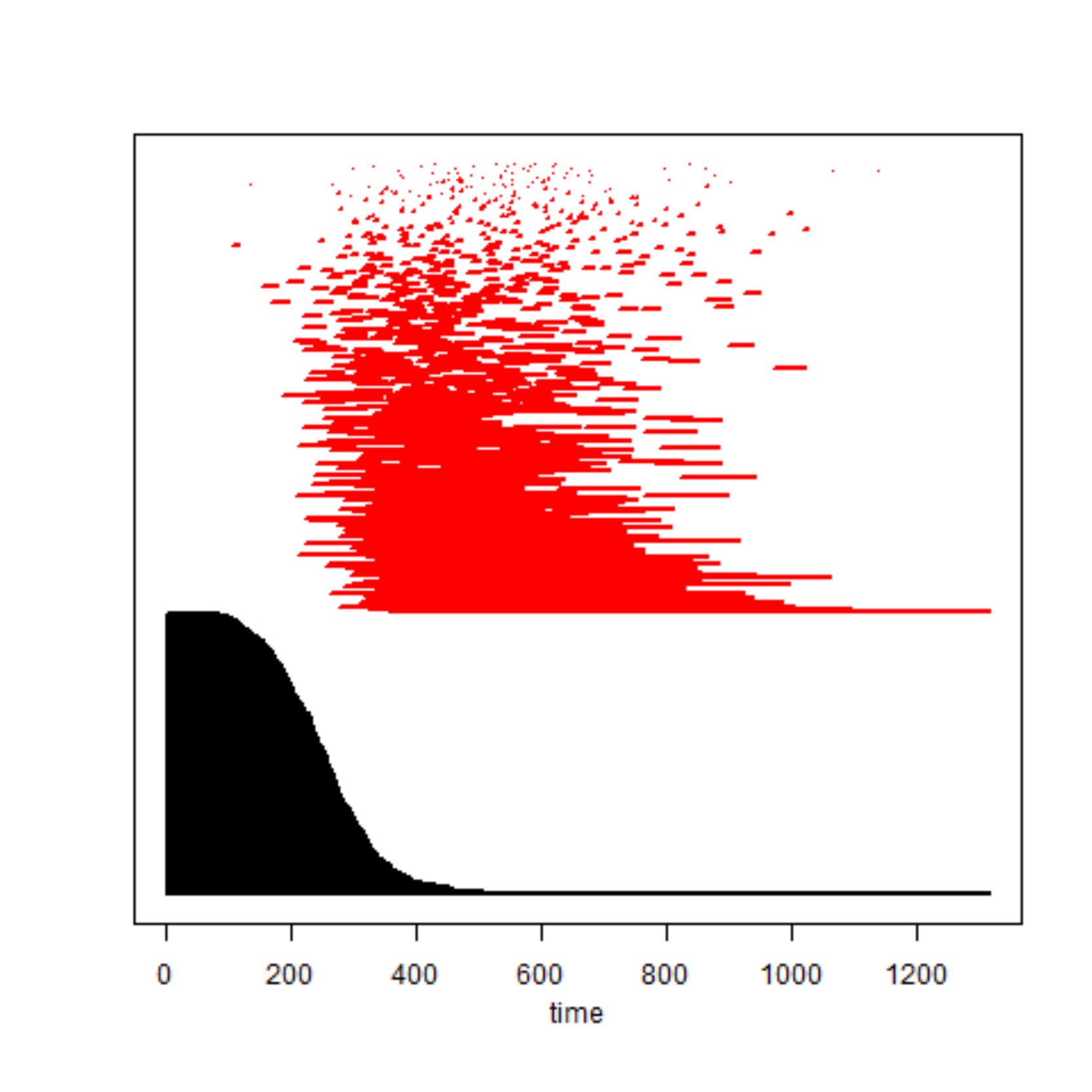}
		\end{tabular}
	\end{center}
	\caption{A dWL image (top-left) and a dWP image (top-right) and their corresponding barcodes at the bottom row.}
	\label{Images}
\end{figure}

After selecting $400$ cells from each image and taking their centroids, we compute their Alpha Complex and persistent homology using the \emph{R} package \cite{TDAR}. See Figure~\ref{Images} for two  examples of processed images and corresponding  barcodes. The later statistical analysis and plots are computed using $R$ as well. Then, we calculate their persistent entropy and display it in Table \ref{tab1}. The whole code used in the process can be found here \url{http://grupo.us.es/cimagroup/downloads.htm}

\begin{table}
	\begin{center}
		\caption{These are the entropy values obtained for each of the barcodes in each image of dimension 0 ($PE_0$), dimension 1 ($PE_1$) and dimension 0 and 1 together ($PE_{all}$)}\label{tab1}
		\resizebox{\columnwidth}{!}{%
		\begin{tabular}{ |c|c|c|c|c|c|c|c|c|}
			\hline
			\multicolumn{3}{|c|}{cNT} & \multicolumn{3}{|c|}{dWL} & \multicolumn{3}{|c|}{dWP} \\
			\hline
			$PE_0$ & $PE_1$ & $PE_{all}$ & $PE_0$ & $PE_1$ & $PE_{all}$ & $PE_0$ & $PE_1$ & $PE_{all}$ \\
			\hline
			\vspace{-0.08cm}
			8.472098 & 8.054819 & 9.271538 & 8.530091 & 8.314505 & 9.405349 & 8.532141 & 8.544773 & 9.485134\\
			\hline
			8.510505 & 8.211333 & 9.354456 & 8.499180 & 8.373499 & 9.405761 & 8.528216 & 8.590107 & 9.497458\\
			\hline
			\vspace{-0.08cm}
			8.501578 & 8.189123 & 9.340218 & 8.474973 & 8.558060 & 9.450642 & 8.565301 & 8.592294 & 9.528683\\
			\hline
			\vspace{-0.08cm}
			8.494586 & 8.224712 & 9.349211 & 8.500253 & 8.321634 & 9.389122 & 8.528192 & 8.113923 & 9.330351\\
			\hline
			\vspace{-0.08cm}
			8.467424 & 8.035914 & 9.262924 & 8.518621 & 8.320296 & 9.397914 & 8.541004 & 8.540452 & 9.493262
\\
			\hline
			\vspace{-0.08cm}
			8.476784 & 8.072562 & 9.280356 & 8.487827 & 8.429975 & 9.416421 & 8.539995 & 8.333815 & 9.421127
\\
			\hline
			\vspace{-0.08cm}
			8.465893 & 8.109965 & 9.287608 & 8.489042 & 8.327057 & 9.382901 & 8.491693 & 8.337785 & 9.389088
\\
			\hline
			\vspace{-0.08cm}
			8.496788 & 8.167088 & 9.327743 & 8.522213 & 8.354212 & 9.413363 & 8.551742 & 8.549792 & 9.501880

\\
			\hline
			\vspace{-0.08cm}
			8.469002 & 8.121784 & 9.294830 & 8.469478 & 8.363328 & 9.382133 & 8.540177 & 8.532990 & 9.490416 \\
			\hline
			\vspace{-0.08cm}
			8.495054 & 8.224863 & 9.347665 & 8.494662 & 8.436416 & 9.423835 & 8.525426 & 8.558831 & 9.490213
\\
			\hline
			\vspace{-0.08cm}
			8.431788 & 7.977363 & 9.216531 & 8.560159 & 8.559574 & 9.507645 & 8.537906 & 8.498011 & 9.474403\\
			\hline
			\vspace{-0.08cm}
			8.491598 & 8.082377 & 9.294716 & 8.474200 & 8.436416 & 9.406070  & 8.552638 & 8.432509 & 9.459623\\
			\hline
			\vspace{-0.08cm}
			8.458547 & 8.169347 & 9.304405 & 8.540980 & 8.468567 & 9.466096 & 8.557452 & 8.404452 & 9.454019\\
			\hline
			\vspace{-0.08cm}
			8.482832 & 8.151599 & 9.314709 & 8.530418 & 8.486440 & 9.463861 & & & \\
			\hline
			\vspace{-0.08cm}
			8.478163 & 8.092100 & 9.289523 & 8.544339 & 8.389995 & 9.440765 & & &\\
			\hline
			8.429276 & 7.992755	& 9.222562 & 		  &	   &      &   & &\\
			\hline
		\end{tabular}
		}
	\end{center}
\end{table}

\begin{figure}
        \centering
            \begin{tabular}{c c}
                \includegraphics[scale=0.4]{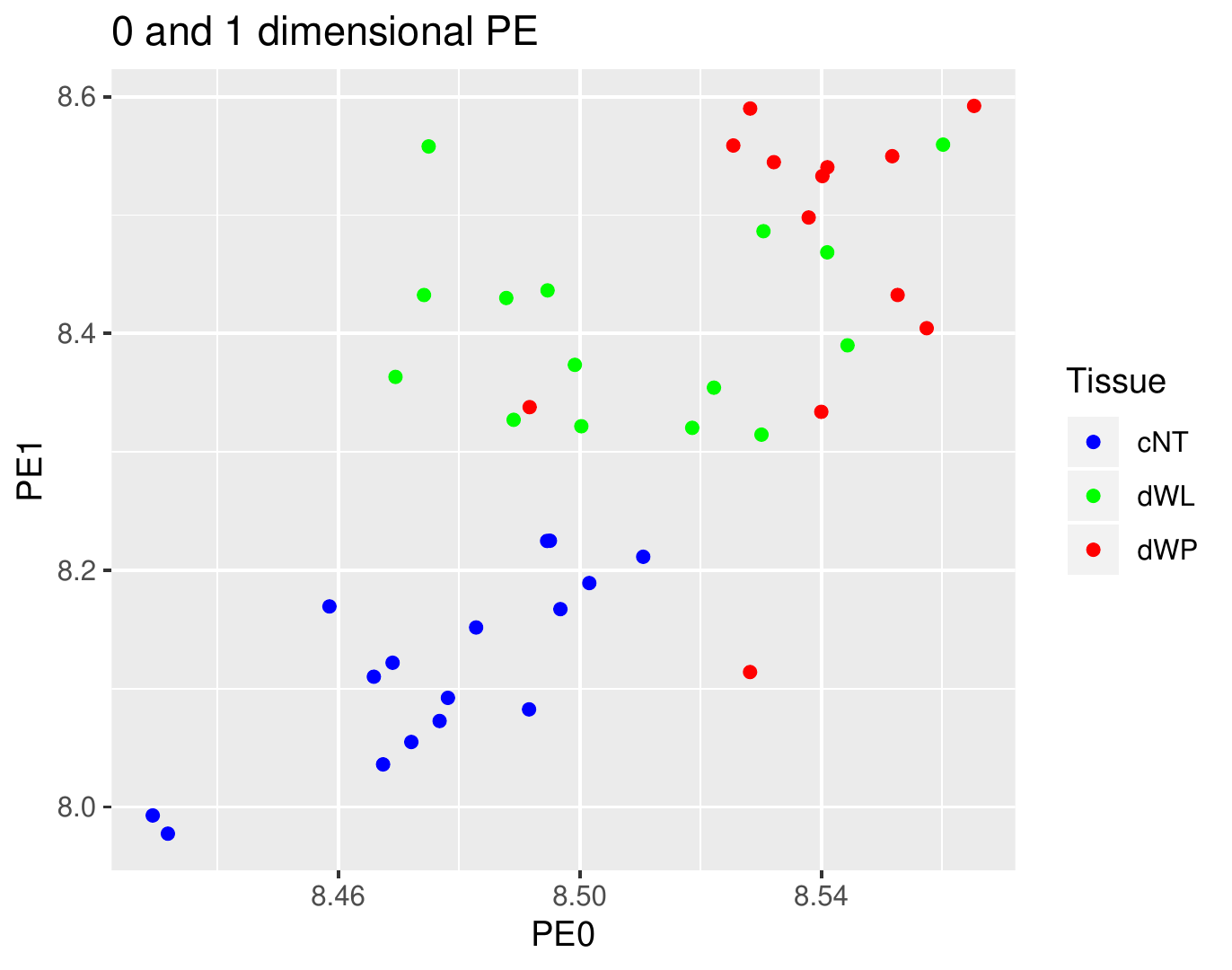}\includegraphics[scale=0.4]{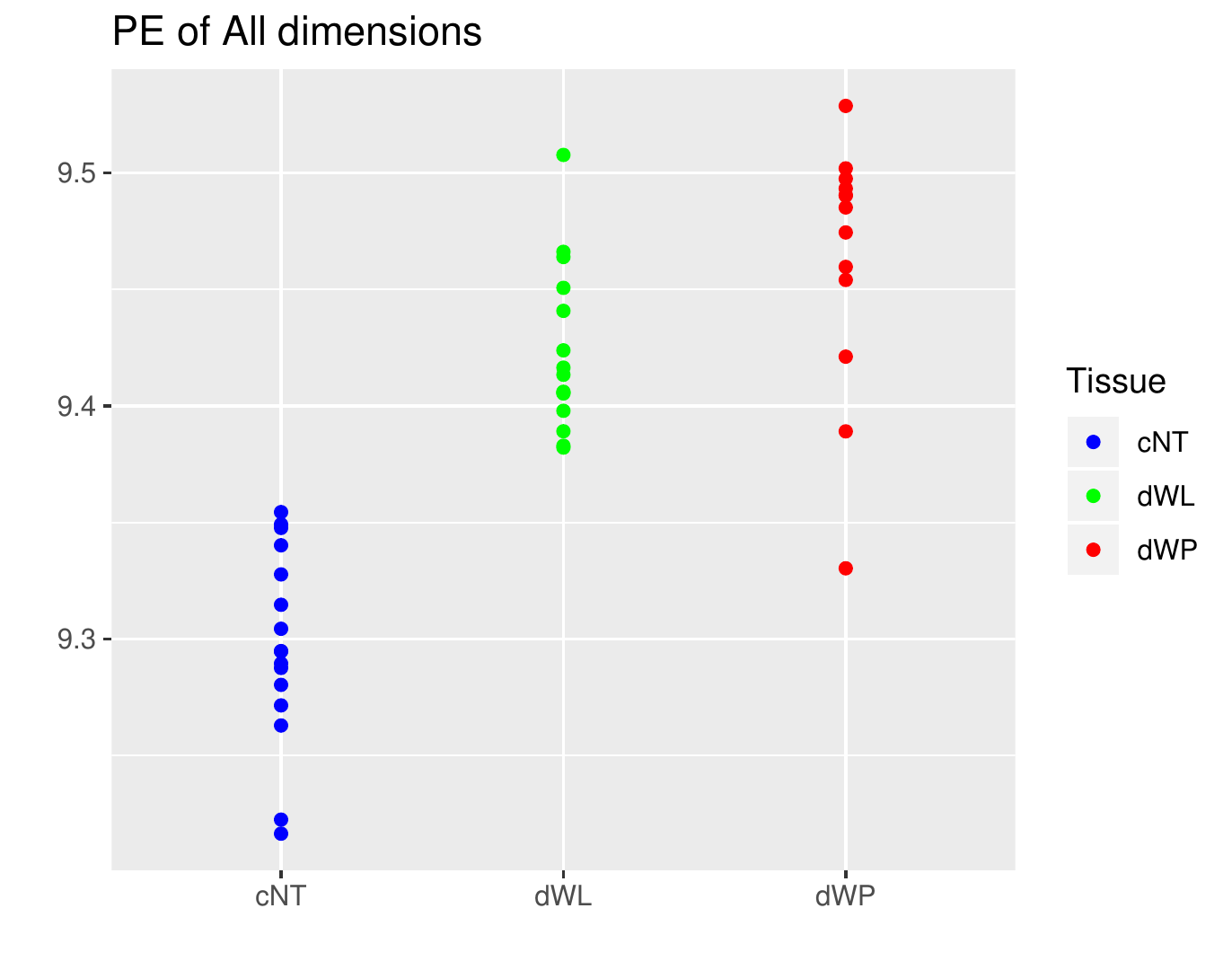}  \\
            \end{tabular}
        \caption{The left image displays $PE_0$ versus $PE_1$ and the right one $PE_{all}$ of the three groups of epithelial tissues.}
        \label{point}
    \end{figure}

\begin{figure}
        \centering
            \begin{tabular}{c}
                \includegraphics[scale=0.5]{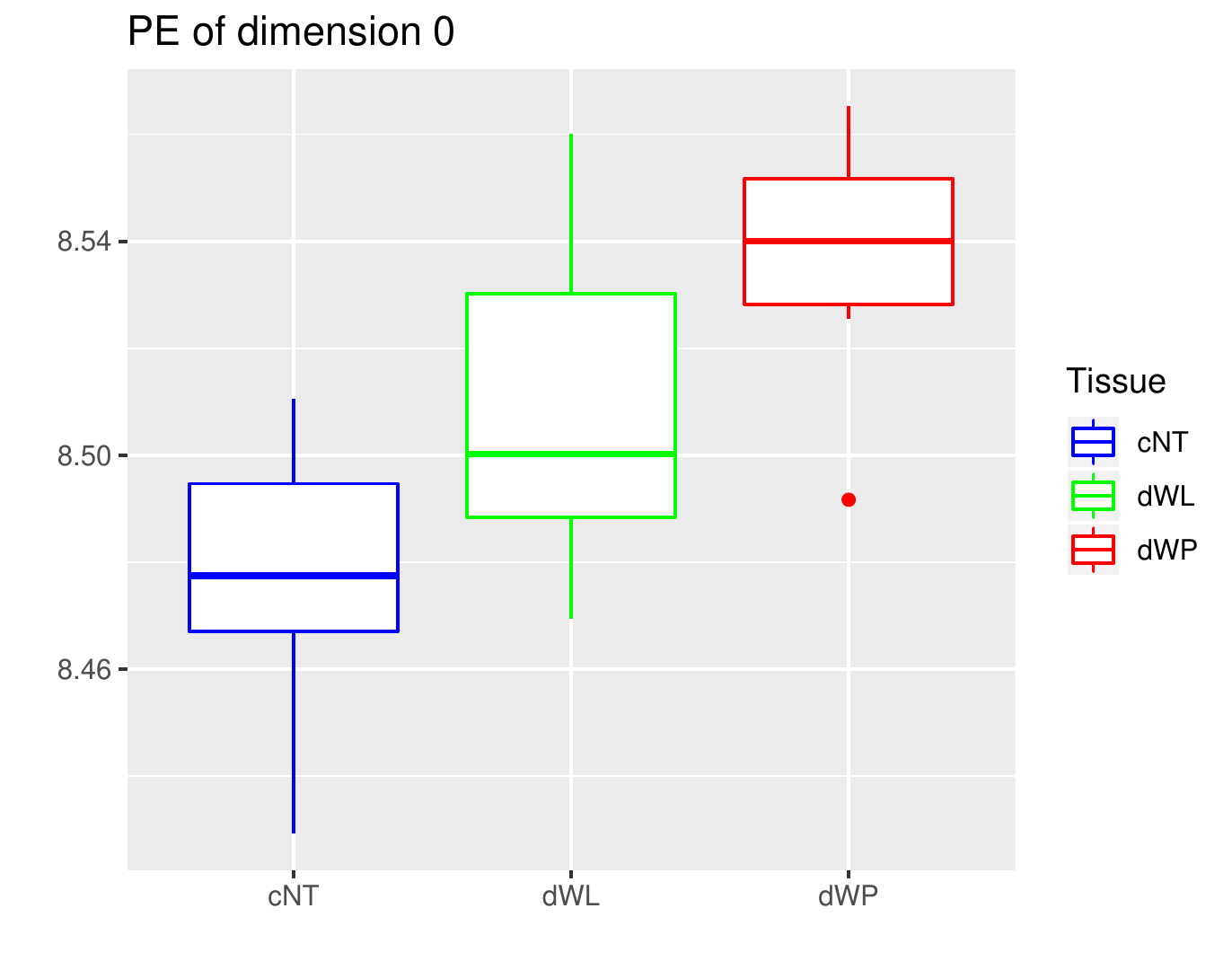} \\ \includegraphics[scale=0.5]{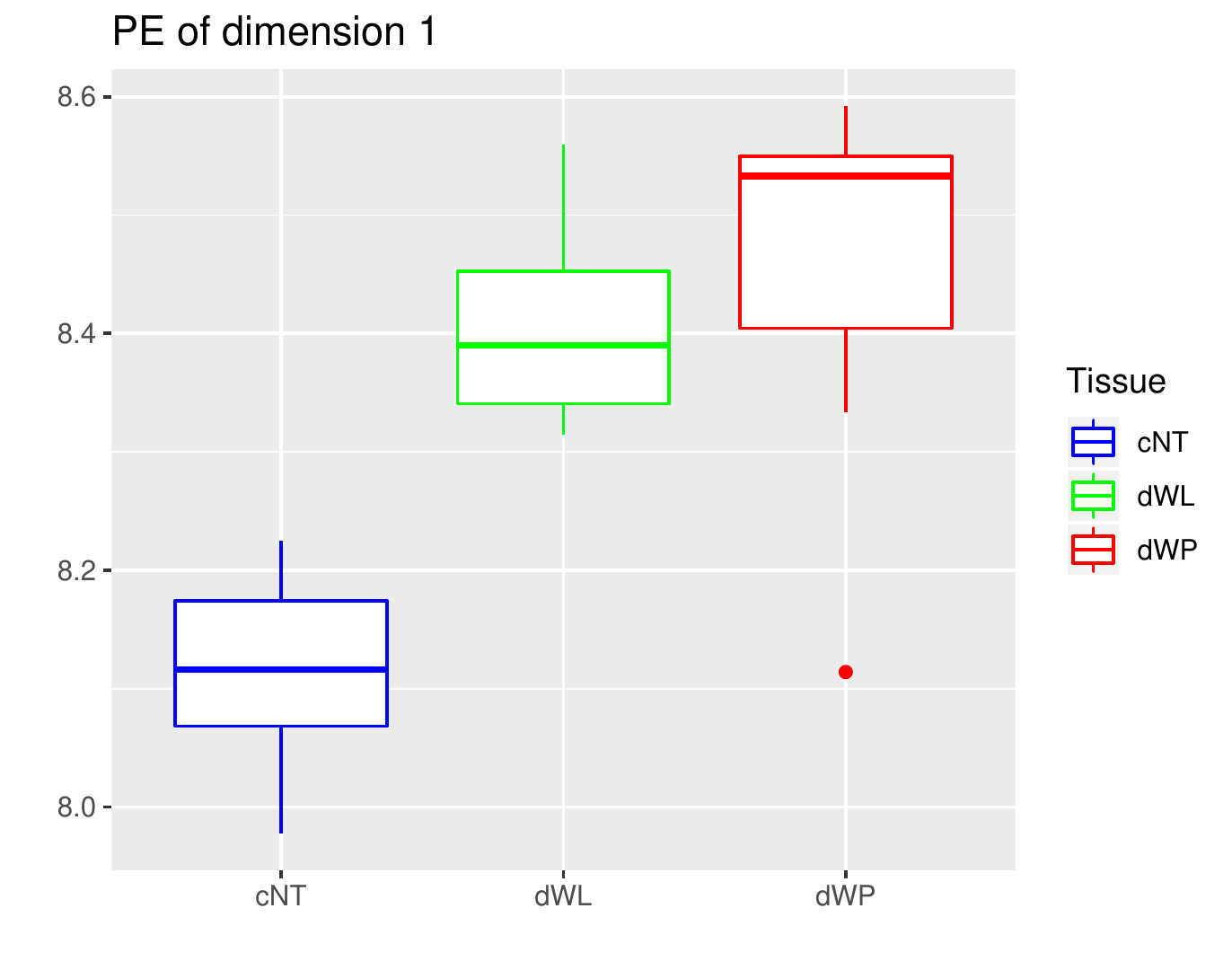} \\
                \includegraphics[scale=0.5]{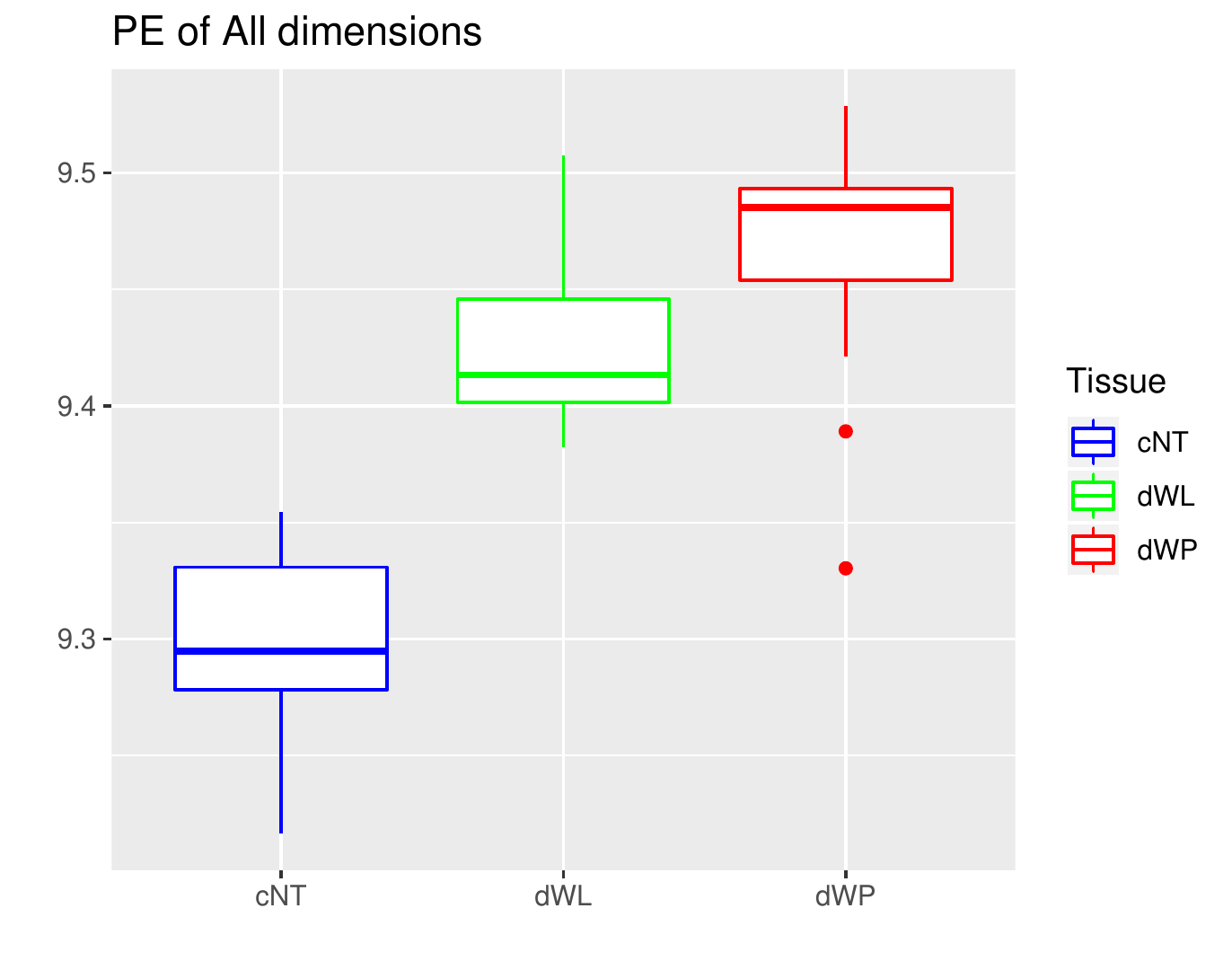}\\
            \end{tabular}
        \caption{From the top to the bottom, the botplox of persistent entropy of dimensions 0, 1 and 0 and 1 together.}
        \label{boxplots}
    \end{figure}

First, we perform a small descriptive statistical study. In Figure \ref{point} we display $PE_0$ versus $PE_1$ in one window and $PE_{all}$ in the other one. Although cNT, dWL and dWP are not perfectly separated, they seem to follow different distributions. These differences are clarified by the boxplots of Figure \ref{boxplots}.

    Thanks to the boxplot representation, it is clear that there may exist differences between the three groups. In order to sustain this idea, we perform the non-parametric multivariate test Kruskal-Wallis to see if there are differences between the three groups simultaneously. After that, we perform a Dunn Test to see the pairwise differences. We will consider that the topology of the cell organization produces different distributions of persistent entropy when the p-value is smaller than $0.005$. Our results are shown in Tables \ref{tab4} and \ref{tab2}. 
    
    $PE_1$ has the best $p$-value when using the Kruskal-Wallis Test although it cannot differentiate dWP and dWL in the Dunn test. The other variables give a small $p$-value as well in Kruskal-Wallis, being $PE_0$ the only one distinguishing all tissues pairwise and $PE_{all}$ the best for separating $cNT$ from the other two.
    
    \begin{table}[]
        \centering
        \caption{Kruskal-Wallis Test for comparing the persistent entropies of the processed tissue images.}\label{tab4}
        \begin{tabular}{|l|l|l|l|l|}
        \hline
        \multicolumn{2}{|l|}{\multirow{2}{*}{\begin{tabular}[c]{@{}l@{}}KWT\\ p-value\end{tabular}}} & $PE_0$    & $PE_1$    & $PE_{all}$ \\ \cline{3-5} 
        \multicolumn{2}{|l|}{}                                                                       & 1.427e-05 & 5.768e-07 & 2.005e-07  \\ \hline
        \end{tabular}
    \end{table}
 
 	 	\begin{table}[]
 		\centering
 		\caption{Dunn Test for comparing the persistent entropies of the processed tissue images.}\label{tab2}
 	\begin{tabular}{|l|l|l|l|l|l|l|l|}
        \hline
        \multicolumn{2}{|l|}{\begin{tabular}[c]{@{}l@{}}DT p-value \\  adjusted\end{tabular}} & \multicolumn{2}{l|}{dWL vs dWP} & \multicolumn{2}{l|}{cNT vs dWL} & \multicolumn{2}{l|}{dWP vs cNT} \\ \hline
        \multicolumn{2}{|l|}{$PE_{0}$}                                               & \multicolumn{2}{l|}{0.02671554}   & \multicolumn{2}{l|}{0.01600541}   & \multicolumn{2}{l|}{7.574294e-06
}  \\ \hline
        \multicolumn{2}{|l|}{$PE_{1}$}                                               & \multicolumn{2}{l|}{0.3271768}    & \multicolumn{2}{l|}{ 5.791831e-05}  & \multicolumn{2}{l|}{2.162007e-06}  \\ \hline
        \multicolumn{2}{|l|}{$PE_{all}$}                                             & \multicolumn{2}{l|}{0.1537159}    & \multicolumn{2}{l|}{1.024837e-04}  & \multicolumn{2}{l|}{3.865447e-07}  \\ \hline
    \end{tabular}
 	\end{table}


\section{Conclusions and Future Work}\label{end}
	We have shown the potential of persistent entropy as a useful topological statistic. In particular, we have applied it to images of three different cellular tissues (cNT, dWL, dWP) to find significant differences between them.\\
	This technique could be improved using a new filtration that used the own regions delimited by the cells instead of approximating them using the Voronoi Diagrams. Also it would be interesting to study more epithelial tissues.\\
	The initial good results presented here may open a door to the inclusion of persistent entropy as one more parameter to be taken into account in analysis tools like \cite{epigraph}.


\bibliographystyle{splncs04}

\end{document}